\immediate\write18{makeindex "root.nlo" -s nomencl.ist -o "root.nls"}
\documentclass[journal]{IEEEtran}
\usepackage{amsfonts,amssymb,amsmath}
\usepackage{algorithm,algorithmic}
\usepackage{booktabs}
\usepackage{graphicx}  
\usepackage{subfigure}
\usepackage{dsfont}
\usepackage{epsfig,epstopdf}
\usepackage{hyperref}
\usepackage{multirow,multicol}
\usepackage{xcolor}
\usepackage{bm}
\usepackage{theorem}
\usepackage{cite}
\usepackage{nomencl}
\usepackage{enumerate}
\usepackage{url}
\makenomenclature
\renewcommand{\nomgroup}[1]{%
\ifthenelse{\equal{#1}{I}}{\item[\textit{Sets}]}{}
\ifthenelse{\equal{#1}{C}}{\item[\textit{Parameters}]}{}
\ifthenelse{\equal{#1}{V}}{\item[\textit{Decision Variables}]}{}
\ifthenelse{\equal{#1}{F}}{\item[\textit{Function}]}{}
\ifthenelse{\equal{#1}{R}}{\item[\textit{Random Variables}]}{}
}
\allowdisplaybreaks[4]

\begin{document}

\title{Federated Reinforcement Learning for Electric Vehicles Charging Control on Distribution Networks}

\author{Junkai Qian,~\IEEEmembership{Graduate Student Member,~IEEE}, Yuning Jiang,~\IEEEmembership{Member,~IEEE}, \\
Xin Liu,~\IEEEmembership{Member,~IEEE}, Qiong Wang, Ting Wang,~\IEEEmembership{Senior Member,~IEEE}, \\
Yuanming Shi,~\IEEEmembership{Senior Member,~IEEE}, and
Wei Chen,~\IEEEmembership{Senior Member,~IEEE}
\thanks{The first two authors contributed equally. This work was supported in part by the Natural Science Foundation of Shanghai under Grant 21ZR1442700, in part by the Shanghai Sailing Program under Grant 22YF1428500, and in part by the Swiss National Science Foundation under
the NCCR Automation (grant agreement 51NF40\_180545). (Corresponding authors: Ting Wang and Xin Liu.)}
\thanks{Junkai Qian and Ting Wang are with the MoE Engineering Research Center of Software/Hardware Co-design Technology and Application, the Shanghai Key Lab. of Trustworthy Computing, East China Normal University, China (e-mail:
{\tt 51255902137@stu.ecnu.edu.cn, twang@sei.ecnu.edu.cn})}
\thanks{Yuning Jiang is with the Automatic Control Laboratory, EPFL, Switzerland (e-mail: {\tt yuning.jiang@ieee.org}).}
\thanks{Xin Liu and Yuanming Shi are with the School of Information Science and Technology, ShanghaiTech University, China (e-mail: {\tt liuxin7, shiym@shanghaitech.edu.cn}).}
\thanks{Qiong Wang is with the State Grid Beijing Electric Power Company, China (email: {\tt wangqiong@bj.sgcc.com.cn})}
\thanks{Wei Chen is with the Department of Electronic Engineering, Tsinghua University, China (e-mail: {\tt wchen@tsinghua.edu.cn}).}}

\maketitle

\begin{abstract}
With the growing popularity of electric vehicles (EVs), maintaining power grid stability has become a significant challenge. To address this issue, EV charging control strategies have been developed to manage the switch between vehicle-to-grid (V2G) and grid-to-vehicle (G2V) modes for EVs. In this context, multi-agent deep reinforcement learning (MADRL) has proven its effectiveness in EV charging control. However, existing MADRL-based approaches fail to consider the natural power flow of EV charging/discharging in the distribution network and ignore driver privacy. To deal with these problems, this paper proposes a novel approach that combines multi-EV charging/discharging with a radial distribution network (RDN) operating under optimal power flow (OPF) to distribute power flow in real time. A mathematical model is developed to describe the RDN load. The EV charging control problem is formulated as a Markov Decision Process (MDP) to find an optimal charging control strategy that balances V2G profits, RDN load, and driver anxiety. To effectively learn the optimal EV charging control strategy, a federated deep reinforcement learning algorithm named FedSAC is further proposed. Comprehensive simulation results demonstrate the effectiveness and superiority of our proposed algorithm in terms of the diversity of the charging control strategy, the power fluctuations on RDN, the convergence efficiency, and the generalization ability.
\end{abstract}


\IEEEpeerreviewmaketitle
\nomenclature[i]{$\mathcal{V}$}{Set of bus in RDN}
\nomenclature[i]{$\mathcal{L}$}{Set of transmission lines in RDN}
\nomenclature[i]{$\varOmega_s$}{Set of children buses of bus $s$ in RDN}
\nomenclature[I]{$\mathcal{Z}_s$}{Set of EVs' indexes connected to bus $s$}

\nomenclature[c]{$p_{s,t}$}{Active power injection at bus s at time slot $t$}
\nomenclature[c]{$q_{s,t}$}{Rective power injection at bus s at time slot $t$}
\nomenclature[c]{$r_{se}$}{Resistance on the line $\left(s,e\right)$}
\nomenclature[c]{$x_{se}$}{Reactance on the line $\left(s,e \right)$}
\nomenclature[c]{$b_{se}$}{Susceptance on the line $\left(s,e\right)$}
\nomenclature[c]{$\underline{v}_{s}/\overline{v}_{s}$}{Lower and upper bound of the square of voltage magnitude at bus $s$}
\nomenclature[c]{$\overline{l}_{se}$}{Upper bound of the square of current magnitude from bus $s$ to $e$}
\nomenclature[c]{$\eta_c/\eta_d$}{Energy transfer efficiency for EV charging and discharging}
\nomenclature[c]{$C_i$}{Battery capacity of EV$i$}
\nomenclature[c]{$\overline{P}_{i}^{\text{G2V}}/\overline{P}_{i}^{\text{V2G}}$}{Upper bound of charging and discharging power for EV$i$}
\nomenclature[c]{$\overline{a}_{i}^{\text{G2V}}/ \overline{a}_{i}^{\text{V2G}}$}{Upper bound of charging and discharging rate for EV$i$}
\nomenclature[c]{$\beta_{i}^{1},\beta_{i}^{2}$}{Parameter of anxiety SoC}
\nomenclature[c]{$\kappa$}{Weight coefficient for balancing TA and RA}
\nomenclature[c]{$\lambda$}{Weight coefficient for power, anxiety and grid reward}
\nomenclature[c]{$\gamma$}{Reward discount parameter}
\nomenclature[c]{$\delta$}{Learning rate for Q network, policy network and temperature coefficient updates}
\nomenclature[c]{$\tau $}{Learning rate for target Q network updates}

\nomenclature[r]{$\xi_t$}{Electricity price at time slot $t$}
\nomenclature[r]{$v_{s,t}$}{Square of voltage magnitude at bus $s$ at time slot $t$}
\nomenclature[r]{$\text{SoC}_{i,t}^{x}$}{Expected SoC for EV$i$ at time slot $t$}
\nomenclature[r]{$\text{SoC}_{i}^{d}$}{Estimation SoC of the coming travel for EV$i$}
\nomenclature[r]{$t_{i}^{a} / t_{i}^{d}$}{Arriving time and departure time of EV$i$}

\nomenclature[v]{$P_{se,t}$}{Active power from bus $s$ to bus $e$ at time slot $t$}
\nomenclature[v]{$Q_{se,t}$}{Reactive power from bus $s$ to bus $e$ at time slot $t$}
\nomenclature[v]{$p_{0,t}$}{Active power obtained from substation at time slot $t$}
\nomenclature[v]{$P_{i,t}$}{Active charging/discharging power of EV$i$ at time slot $t$}
\nomenclature[v]{$P_{agg_s,t}$}{Active power on aggregator $s$ at time slot $t$}
\nomenclature[v]{$a_{i,t}$}{Charging/discharging rate for EV$i$ at time slot $t$}
\nomenclature[v]{$\text{SoC}_{i,t}$}{State of charge for EV$i$}
\nomenclature[v]{$s_{i,t}$}{State of EV$i$ at time slot $t$}
\nomenclature[v]{$\theta^{k}/\hat{\theta}^{k}$}{Parameter of critic and target critic network}
\nomenclature[v]{$\phi$}{Parameter of actor network}
\nomenclature[v]{$\alpha$}{Temperature coefficient}
\nomenclature[v]{$\epsilon $}{Reparameterization parameter for policy network}
\nomenclature[v]{$\sigma_g $}{Standard deviation of power change in the test}

\printnomenclature[0.55in]

\section{Introduction}

\IEEEPARstart{E}{lectric} vehicles (EVs) offer a promising alternative to conventional fossil fuel-powered vehicles, with their potential to mitigate greenhouse gas (GHG) emissions and air pollution \cite{zhang2018comparison}. 
Recent projections from the International Energy Agency (IEA) indicate that the EV market share has substantially increased in recent years and is predicted to reach approximately 30\% by 2030 \cite{iea2021global}. However, the increase in EVs also poses new challenges for the power grid, including burdening the electricity loads and amplifying the peak electricity demands~\cite{khalid2019comprehensive, nour2020review, jiang2021data}. \cite{10EVs} demonstrated that even a 10\% increase in EVs could result in significant fluctuations in the grid's voltage curves.

Vehicle-to-grid (V2G) and grid-to-vehicle (G2V) are considered effective techniques to minimize the negative impact of EV charging on the grid \cite{V2G, sovacool2020actors}. 
V2G involves discharging an EV back into the power grid, while G2V refers to charging an EV from the grid. 
The V2G/G2V mode in a distribution network can help stabilize the power curve \cite{lin2014optimal, sun2019optimized, chen2021multimicrogrid} and provide peak shaving/valley filling services \cite{tang2018distributed}. 
However, the uncertain factors, including dynamic electricity prices, grid load, and uncertainty in human behavior, increase the difficulty of generating a generalized charging policy \cite{wang2016predictive, muratori2018impact, jin2022shortest}. Consequently, the critical challenge lies in effectively coordinating EVs' charging and discharging behavior in a dynamic environment to enable their active participation in electricity delivery \cite{moschella2021decentralized}.


Numerous approaches have been proposed to develop optimal EV charging control policies. Traditional studies have formulated charging control as an optimization problem that aims to maximize the drivers' profits \cite{cao2021smart}. For instance, the work \cite{two-stage} proposes a two-stage charging optimization model to reduce the operation cost at the workplace. In \cite{lp1}, the capacity limitation in distribution transformers is considered, and the continuous EV charging is formulated as linear programming. The authors of \cite{mixed-integer-lp} investigate energy trading in coordination with V2G and propose a solution using mixed-integer stochastic linear programming. \cite{dp1} designs a cost function with battery energy constraints, and the optimal EV charging control is computed through dynamic programming. The work \cite{MDP} introduces the uncertainty in wind energy supply and formulates the EV charging control as a Markov Decision Process (MDP). Reference \cite{chung2020intelligent} considers the customers' dynamic preference for charging parameters and conducts a stochastic game to address the uncertainty. Reference \cite{timeanxiety} introduces time anxieties to address uncertain events and solves the EV charging problem using a generalized Nash equilibrium (NE) game.
In summary, all the algorithms mentioned above maintain an explicit optimization model under the assumption of a fully observable environment called model-based algorithms. The optimal charging control strategy is generated based on an accurate system model. However, fully observable environments rely on a priori knowledge about state transitions, and uncertain factors in the environment can interfere with state transitions, which further affects the model's accuracy.

Model-free deep reinforcement algorithm (DRL) has demonstrated great potential in sequential decision-making problems without prior knowledge of the environment's dynamics \cite{RL-review1, RL-review2,RL-review3}. In DRL, the agent makes a decision for each state to obtain the reward sequence and then learns to optimize its policy to maximize expected future rewards. DRL algorithms have been widely adopted in grid management and have achieved excellent performance \cite{RL-grid1}, including in EV charging control. 
In \cite{wan2018model}, discriminative features of electricity prices are extracted through a representation network, and Q-learning is developed to obtain the optimal discrete EV charging control strategy. 
Reference \cite{zhang2020cddpg} utilizes the long short-term memory (LSTM) method to extract the trend of electricity prices. The estimated electricity price is applied in a deep deterministic policy gradient (DDPG) algorithm for continuous EV charging control policy. In \cite{qian2019deep}, the traffic conditions are considered to minimize the travel time and charging costs during charging navigation. \cite{li2019constrained} formulates the EV charging control problem as a constrained MDP and develops a safe DRL method to learn the optimal scheduling strategy. 
In \cite{yanlingfang}, the authors mathematically model the driver's anxiety using statistical principles and propose using a soft actor-critic (SAC) to learn the optimal policy. 
However, EV charging behaviors in the same system tend to affect each other, where such mutual influence is often ignored in single-agent DRL-based approaches.
Multi-agent deep reinforcement learning (MADRL) algorithms are more suitable for addressing EV charging control problems than single-agent ones \cite{mai2020multi}. For example, the work \cite{da2019coordination} promotes the type of tariff and develops a multi-agent multi-objective reinforcement learning framework to optimize the charging schedule. In \cite{jin2020optimal}, a multi-agent SAC-based framework is proposed to learn the environment's dynamics, and its performance is verified in a simulated distribution network. In \cite{yan2022cooperative}, the authors study the transformer constraints in a residential system and approximate the collective behaviors of EVs through a collective-policy model, using multi-agent SAC to learn the charging scheduling strategy. 
Overall, model-free DRL-based methods have proven to be superior in addressing uncertainty in EV charging control.

Nevertheless, the approaches mentioned above have two drawbacks. Firstly, while some advanced methods (e.g., \cite{lp1,jin2020optimal,yan2022cooperative}) combine EV charging control with the distribution network and consider a systematic optimization process, they usually use simplified constraints as a substitute for complex power flow in the distribution network, which is insufficient and unreliable. 
Moreover, the substitute electricity constraints are considered an optimization objective, which is often optional, leading to the possibility of irreversible damage to the distribution network during the EV charging control. 
Secondly, the charging and travel data of each EV is privacy-sensitive, raising concerns about driver privacy when sharing massive data in MADRL.
To address this, some latest MADRL algorithms (e.g., \cite{yan2022cooperative}) maintain a charging control strategy with individual charging experience to avoid privacy breaches. However, the isolated data of each EV limit the generalization of the strategy \cite{shi2023vertical, letaief2021edge, yang2022lead}.

To tackle the above issues, we propose an approach that leverages optimal power flow (OPF) \cite{frank2012optimal} to model radial distribution networks (RDNs) \cite{chakravorty2001voltage} and federated learning (FL) \cite{zhan2020learning, kairouz2021advances, wang2021federated, yang2020federated, qi2021federated}  for charging control policy training. 
Specifically, we study a multi-EV charging and discharging system on an RDN, which operates under OPF and aims to allocate power flows to minimize power loss. Each EV is controlled by an agent that deploys a soft actor-critic (SAC) model inspired by \cite{yanlingfang}. SAC has achieved excellent effects in continuous action space. It uses an entropy regularization method, which can effectively control the exploration and stability of the policy. Each agent receives regulation information from the RDN and executes local model training to maximize rewards, including V2G profits, grid idle degree, and driver satisfaction. The training process follows the FL mechanism, where agents upload their local model parameters to the server via a communication link. The server aggregates these parameters to refine a global model that enhances the generalization of the charging control strategy. As only model parameters are transferred during transmission, and local data is stored on each EV, drivers' privacy is guaranteed \cite{shi2020communication, shi2023task, yang2021privacy}.

To summarize, this paper makes the following main contributions:
\begin{itemize}
\item We introduce an RDN that operates under OPF to combine with a multi-EV charging control system. By considering electric component constraints, we simulate the EV charging/discharging power flow on the RDN, resulting in more solid and realistic constraints than quantitative approaches.

\item We develop a grid reward with OPF-based RDN, introducing a new optimization objective, i.e., grid idle degree, in the case of unbalanced charging and discharging rates. The EV agents trained through the grid reward will sequentially charge/discharge when the grid is idle, significantly reducing the power fluctuations on the RDN.

\item We model the EV charging control problem as a multi-agent MDP and propose a continuous federated SAC algorithm called FedSAC. Compared to previous standard DRL methods, our FedSAC approach achieves the best control performance.
\end{itemize}


The rest of this paper is organized as follows. Section \ref{sec::model} presents the EV charging/discharging model on an OPF-based RDN and models the 
EV charging/discharging process as an MDP. Section \ref{sec::algorithm} introduces the overall framework of the proposed FedSAC. In Section \ref{sec::experiment}, simulations are conducted to verify the performance of the proposed algorithm. Section \ref{sec::conclusion} concludes the paper.

\section{The Distribution Network Multi-EV G2V/V2G Model}
\label{sec::model}

This paper considers the radial distribution network (RDN) integrated multi-EV G2V/V2G model. Each EV is equipped with a G2V/V2G management module, which works as an agent to conduct model training and control the charging process. In this section, we first combine EV charging/discharging with a radial distribution network and introduce OPF to model the system. Then, the driver's charging anxiety is modeled to consider the different habits of the driver. Finally, the charging control process is formulated as an MDP with hourly dynamic electricity prices.

\subsection{The Radial Distribution Network Model}

The radial distribution network refers to the power network that receives electricity from the transmission network or regional power plants. Through distribution facilities, it distributes electricity to various users regionally or level by level according to voltage. 
\begin{figure*}[htbp!]
\centering
\includegraphics[width=.95\linewidth]{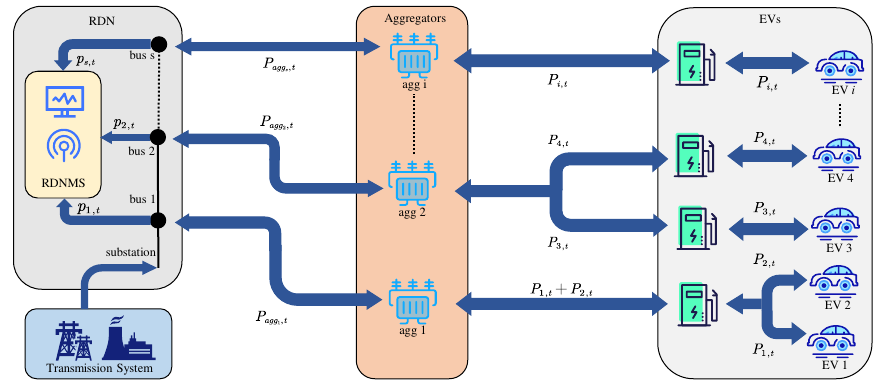}
\caption{The structure of EVs connected to RDN}
\label{model structure}
\end{figure*}

As discussed in \cite[Section II.A]{powerconstraints}, we consider the AC power flow on RDN. To this end, we define the topology of RDN by $(\mathcal{V},\mathcal{L})$, where $\mathcal{V}:=\left\{ 0,\cdots,v \right\}$ denotes the set of buses. Without loss of generality, the substation is denoted as bus $0$, $\mathcal L\subseteq \mathcal V\times \mathcal V$ denotes the set of transmission lines. Then, let $p_{s,t}$ and $q_{s,t}$ denote the active and reactive power injection at bus $s$ at time slot $t$.  For the transmission line $(s,e)\in\mathcal L$,
we define by $r_{se}$ and $x_{se}$ the resistance and the reactance, respectively and by $b_{se}$ the total charging susceptance. Moreover, $P_{se,t}$ and $Q_{se,t}$ denote the active and reactive power from bus $s$ to $e$ during $[t,t+1)$, respectively. Thus, the AC power equation in a radial distribution network can be described as
\begin{subequations}
\label{AC power equation}
\begin{align}
P_{se,t} &=p_{s,t} +P_{agg_s,t} +\sum_{h\in \varOmega _s}{\left( P_{hs,t} -r_{hs} l_{hs,t} \right)}, \\
Q_{se,t} &=q_{s,t} +\sum_{h\in \varOmega _s}{\left( Q_{hs,t} -x_{hs} l_{hs,t} \right)},
\end{align}
\end{subequations}
where $\varOmega_s$ denotes the index set of children neighbors of bus $s$, and $P_{agg_{s},t}$ is the charging power of the aggregator connected to bus $s$ during $[t,t+1)$. Moreover, $l_{hs}$ is the square of current magnitude, i.e.,
\begin{equation}
l_{se,t}=\frac{ P_{se,t} ^2+ Q_{se,t}^2}{v_{s,t}},
\end{equation}
where $v_{s,t} \in [\underline{v}_{s},\overline{v}_{s}]$ is the square of voltage magnitude at bus $s$ during $[t,t+1)$. Then, the equations of power flow from $s$ to $e$ are represented as
\begin{equation}
v_{s,t}-v_{e,t}=2\left( r_{se}P_{se,t}+x_{se}Q_{se,t} \right) -l_{se,t}\left(r_{se}^2+ x_{se}^2 \right).
\end{equation}
In addition, flows in the branch's charging susceptance should also be considered. Thus, the capacity limits on all branches during $[t,t+1)$ are defined as 
\begin{subequations}
\begin{align}
l_{se,t}+\frac{1}{4}v_{s,t} b_{se} ^2+b_{se}Q_{se,t}& \le \overline{l}_{se}, \\
l_{se,t}+\frac{1}{4}v_{e,t}b_{se}^2+b_{se}\left( x_{se}l_{se,t}-Q_{se,t} \right)& \le \overline{l}_{se}.
\end{align}
\end{subequations}

\subsection{EV Charging Model on RDN}

Let us consider $N$ EVs with the same battery and charging infrastructures. Each EV can work in both G2V and V2G modes. We assume that EV$i$ connects the charging pile at time $t_{i}^{a}$ and departs at time $t_{i}^{d}$. The State of Charge (SoC) of EV$i$ at time $t$ is denoted as $\text{SoC}_{i,t}$, which satisfies $\text{SoC}_{i,t}\in \left[ 0,1 \right]$. The charging and discharging process of EV$i$'s battery at time slot $t$ is modeled as
\begin{equation}
\text{SoC}_{i,t+1}=\begin{cases}
\text{SoC}_{i,t}&\quad t<t_{i}^{a},\ t\ge t_{i}^{d},\\
\text{SoC}_{i,t}+a_{i,t}&\quad t_{i}^{a}\le t<t_{i}^{d},\\
\end{cases}
\end{equation}
where $a_{i,t}$ is the continuous charging/discharging rate at time slot $t$, i.e., $a_{i,t}>0$ when EV$i$ work in G2V mode and $a_{i,t}<0$ when EV$i$ work in V2G mode.

The actual power EV$i$ absorbed/released from RDN \cite{jin2020optimal} during $[t,t+1)$ is calculated as
\begin{equation}
\label{actual power of EV}
\begin{aligned}
P_{i,t}  =&\begin{cases}
P_{i,t}^{\text{G2V}}& a_{i,t}>0\\
-P_{i,t}^{\text{V2G}}&  a_{i,t}\leq 0\\
\end{cases}  \\
=&\begin{cases} 
\frac{a_{i,t}\cdot \text{C}_i}{\eta _c}& a_{i,t}>0\\
-\eta _da_{i,t}\text{C}_i& a_{i,t}\leq 0 ,\\
\end{cases}
\end{aligned}
\end{equation}
where $\eta_c(\eta_d)$ is the (dis-)charging efficiency, and $\text{C}_i$ denotes the actual battery capacity of EV$i$.

The aggregator is a middleware between RDN and EVs responsible for receiving charging/discharging power and transmitting control signals. Thus, the power of the aggregator $s$ is equal to the sum of EVs managed by it:
\begin{equation}
\label{agg consists of EV}
\begin{aligned}
P_{agg_s,t}=&\sum_{i\in \mathcal{Z}_s}{P_{i,t}}\\
=&\sum_{i\in \mathcal{Z}_s}P_{i,t}^{\text{G2V}}-\sum_{i\in \mathcal{Z}_s}P_{i,t}^{\text{V2G}}\\
=&\;P_{agg_s,t}^\mathrm{G2V} -P_{agg_s,t}^\mathrm{V2G} ,    
\end{aligned}
\end{equation}
where $\mathcal{Z}_s$ denotes the set of EVs' indexes connected to bus $s$ and $P_{agg_s,t}^{\text{G2V}}$/$P_{agg_s,t}^{V2G}$ denotes the total charging/discharging rate of EVs on the aggregator $s$. Thus, $P_{agg_s,t}$ can be calculated by the charging and discharging rate of EVs connected to aggregator $s$. Additionally, the constraint of charging/discharging power is denoted as 
\begin{equation}
\label{constraint of charging/discharging power}
-\overline{P}_{i}^{\text{V2G}} \le P_{i,t} \le \overline{P}_{i}^{\text{G2V}}, 
\end{equation}
where the $\overline{P}_{i}^{\text{V2G}}$ and $\overline{P}_{i}^{\text{G2V}}$ are the maximum discharging and charging power, respectively. 

\subsection{OPF Formulation on RDN}
When the structural parameters and load conditions of RDN are given, the objective of the OPF is to minimize power loss by rationally distributing the power flow. The power loss at time $t$ is defined as
\begin{equation}
\notag
P_{t}^{\text{loss}}=P_{t}^{\text{sub}}+P_{t}^{\text{V2G}}-P_{t}^{\text{G2V}}-P_{t}^{\text{others}} ,
\end{equation}
where $P_{t}^{\text{sub}}$ denotes the power obtained from substation and $P_{t}^{\text{others}}$ denotes the other load on RDN. $P_{t}^{\text{V2G}}$ denotes the power absorbed from the discharging EVs, and $P_{t}^{\text{G2V}}$ denotes the power cost by the charging EVs, i.e.,
\begin{align}
P_{t}^{\text{V2G}}=\sum_{s \in \mathcal{V}}{P_{agg_s,t}^\mathrm{V2G}}, \;\; P_{t}^{\text{G2V}}=\sum_{s \in \mathcal{V}}{P_{agg_s,t}^\mathrm{G2V}}. \notag
\end{align}

In the article, we assume the other load $P_{t}^{\text{others}}$ remains constant over time for simplification. 
The charging/discharging power $P_{t}^{\text{G2V}}$ and $P_{t}^{\text{V2G}}$ are managed by individual EVs. They are fixed for RDN at every timeslot. Therefore, for optimization problems in RDN, the $P_{t}^{\text{G2V}}$ and $P_{t}^{\text{V2G}}$ can be treated as parametric input. 
In other words, the OPF problem is equivalent to minimizing the energy obtained from the substation at time $t$. The objective function to be optimized can be written as
\begin{subequations}
\label{OPF problem}
\begin{align}\label{OPF obj}
\text{minimize}\;\; &f_{\text{obj},t}(P_{agg,t}):=P_t^\mathrm{loss}\\
\text{subject to}\;\;&P_{agg,t}\in \mathcal{F}_t:=\left\{\eqref{AC power equation} -\eqref{constraint of charging/discharging power} \text{ are satisfied}\right\},
\end{align}
\end{subequations}
where $P_{agg,t}$ stacks $P_{agg_s,t}$ for all $s\in\mathcal V$. We assume the RDN works under OPF all the time in the paper. 

In the RDN, the management system is responsible for power flow distribution with OPF and sends the results to EVs through the communication unit. A charging pile can usually provide charging services for multiple EVs. Fig. \ref{model structure} shows the structure of the RDN. The aggregator acts as a middleware between RDN and EVs, collecting EVs' power and working as a communication link. The aggregators are connected to the lower V2G charging piles, which can provide charging services for multiple EVs.

\subsection{Charging Anxiety Model}

This paper further considers the driver's charging anxiety, including the driver's range anxiety (RA) and time anxiety (TA). 
RA represents the driver's anxiety of depleting the battery, which can be relieved by a higher estimated SoC level at time $t_{i}^{d}$. However, a higher SoC at $t_{i}^{d}$ will also reduce profits obtained from the V2G process. The accuracy of the driver's SoC estimation increases with their experience. TA represents the driver's anxiety about uncertain factors, which may interrupt the charging process and thus leave insufficient power for upcoming travel. Different drivers have different tolerance for uncertain factors.

The charging anxiety is defined by the following mathematical model as presented in \cite{yanlingfang}, which maps the driver's anxiety to the expected SoC:
\begin{equation}
\label{sox}
\text{SoC}_{i,t}^{x}=\frac{\beta_{1,i}\left( e^{{-\beta_{2,i}\left( t-t_{a,i} \right)}/{\left( t_{d,i}-t_{a,i} \right)}}-1 \right)}{e^{-\beta_{2,i}}-1} 
\end{equation}
for $[t_{i}^{x},t_{i}^{d})$, where $t_{i}^{x}\in[t_{i}^{a},t_{i}^{d})$ is the time when the driver begins to feel anxious, hyperparameters $\beta_{i}^{1}\in [0,1]$ and $\beta_{i}^{2}\in \left( -\infty,0 \right) \cup \left( 0, \infty \right)$ determines the growth trend of anxiety SoC. Specifically, larger $\beta_{i}^{1}$ determines higher SoC at $t_{i}^{d}$ while larger $\beta_{i}^{2}$ leads to reaching the expected SoC earlier. 

\subsection{Markov Decision Process Formulation}
The EV charging control process is modeled as a Markov decision process (MDP), which considers the diverse behavior of different EV drivers. For agent $i$, the individual MDP model is formalized by a tuple $\mathcal{M}_i=\left<\mathcal{S}_i,\mathcal{A}_i,\mathbb{P}_i,\mathcal{R}_i,\gamma \right>$, where $\mathcal{S}_i$ represents the state set, $\mathcal{A}_i$ represents the action set, $\mathbb{P}_i$ represents the state transition probability, $\mathcal{R}_i$ represents the reward set, and $\gamma$ represents the discount factor.

\subsubsection{State} 
For agent $i$, the state gained from the environment is utilized as input of charging control policy to generate current action. Specifically, the state at time t is defined as
\[
s_{i,t}=\left( \xi _{i,t-n+1},\cdots ,\xi _{i,t},t_{i}^{d},t_{i}^{x},\text{SoC}_{i,t},\text{SoC}_{i,t}^{x},\text{SoC}_{i}^{d} \right),
\]
where $\left( \xi _{i,t-n+1},\xi _{i,t-n+2},\cdots ,\xi _{i,t} \right)$ is the hourly electricity price of past $n$ hours, $\text{SoC}_{i,t}$ is the EV$i$'s current SoC and $\text{SoC}_{i}^{d}$ equal to $\beta_{i}^{1}$ is the expected SoC when EV$i$ leaves. 
Each charging station has an independent pricing right, so each EV faces different electricity prices simultaneously. However, the pricing of charging stations depends on the procurement cost of electricity, which follows certain laws. Thus, past electricity prices are used to summarize the trend of electricity prices.
The $\text{SoC}_{i}^{d}$ depends on the driver's estimation of the coming travel, and the $\text{SoC}_{i,t}^{x}$ shown in \eqref{sox} represents the drivers expected SoC at time $t$.
The other variables are related to drivers' different charging and travel habits. The variables $t_{i}^{x}$ and $\text{SoC}_{i,t}^{x}$ reflect the $i$-th driver's anxiety trend over time. Drivers with higher anxiety expect the battery to maintain higher SoC during charging. The variables $t_{i}^{d}$ and $\text{SoC}_{i}^{d}$ depend on the driver's estimation of the upcoming travel plan. 

\subsubsection{Action}
The action $a_{i,t}$ denotes the charging/discharging rate of EV$i$ at time slot $t$ with the given $s_{i,t}$. The charging rate $a_{i,t}$ has continuous action space, which is restricted by
\[    
-\overline{a}_{i}^{\text{V2G}} \leq a_{i,t} \leq \overline{a}_{i}^{\text{G2V}} .
\]
Here, $\overline{a}_{i}^{\text{G2V}}$ and $\overline{a}_{i}^{\text{V2G}}$ obtained from \eqref{actual power of EV} and \eqref{agg consists of EV} denote the maximum charging and discharging rate respectively. $a_{i,t}>0$ when EV$i$ works in G2V mode and $a_{i,t}<0$ when EV$i$ works in V2G mode. In our formulation, the maximum charging rate $\overline{a}_{i}^{\text{G2V}}$ and discharging rate $\overline{a}_{i}^{\text{V2G}}$ are modeled as an imbalance to suit reality \cite{V2G,ehsani2018modern}. We assume all EVs have the same battery and charging infrastructures.

\subsubsection{State Transition}
The behaviors of different drivers and random factors result in different environments for each EV in our MDP formulation. Each EV$i$ thus has different state transition functions $\mathbb{P}_i$ affected by personalized drivers' behavior, dynamic electricity price changes, and random factors. We use notation $\mathbb{P}_i\left( s_{i,t+1}|s_{i,t},a_{i,t} \right) $ to denote the probability of transition from state $s_{i,t}$ to $s_{i,t+1}$ when $a_{i,t}$ is taken.

\subsubsection{Reward}
Compared to~\cite{yanlingfang}, we design a reward $r_{i,t}$ such that the impact of EV charging/discharging on the RDN can be considered. Besides, it also considers dynamic electricity prices and drivers' anxiety. The reward is obtained from the local environment unique to each EV and the global environment shared by all EVs. The local environment is an individual EV with a G2V/V2G management module. The global environment is the RDN comprising various electrical components and microcomputers.

The G2V costs and V2G profits are calculated in the local environment with electricity price and charging rate. Thus we define the power reward of EV$i$ at time $t$ as 
\begin{equation}
\label{power reward}
r_{i,t}^{\text{p}}=-\xi _t\cdot a_{i,t},\;\; t_{i}^{a}\le t<t_{i}^{d}. 
\end{equation}
The mathematical model in \eqref{sox} maps the driver’s anxiety to the expected SoC. Additional costs are spent on the amount of charged power to satisfy the driver's anxiety. The time anxiety reward is calculated in the local environment, i.e.,
\begin{equation}
r_{i,t}^{\text{ta}}=\begin{cases}
0& t_{i}^{a}\le t<t_{i}^{x} , \\
-\max \left( \text{SoC}_{i,t}^{x}-\text{SoC}_{i,t},0 \right)^2 & t_{i}^{x}\le t<t_{i}^{d} . \\
\end{cases} 
\end{equation}
When $t_{i}^{a}\le t<t_{i}^{x}$, the driver has not started to feel anxious about the battery level. No additional power is charged, so the cost is zero. When $t_{i}^{x}\le t<t_{i}^{d}$, the term $-\max \left( \text{SoC}_{i,t}^{x}-\text{SoC}_{i,t},0 \right)^2$ denotes a penalty to additional charged power in satisfaction of TA. The penalty term is made nonlinear by using the square function. An intuitive statement is that penalty increases at an accelerating rate when the difference between $\text{SoC}_{i,t}^{x}$ and $\text{SoC}_{i,t}$ increases.

The range anxiety reward is calculated locally as
\begin{equation}
\label{anxiety reward}
r_{i,t}^{\text{ra}}=-\max \left( \text{SoC}_{i}^{d}-\text{SoC}_{i,t},0 \right)^2,\;\; t=t_{i}^{d}.
\end{equation}
At departure time $t_{i}^{d}$, the term $-\max \left( \text{SoC}_{i}^{d}-\text{SoC}_{i,t},0 \right)^2$ is similary a penalty to RA. We combine time anxiety reward and range anxiety reward to obtain the anxiety reward:
\begin{equation}
r_{i,t}^{\text{a}}=\kappa _{\text{ta}}r_{i,t}^{\text{ta}}+\kappa _{\text{ra}}r_{i,t}^{\text{ra}},
\end{equation}
where $\kappa _{\text{ta}}$ and $\kappa _{\text{ra}}$ control sensitivity to different anxieties according to the driver's habit. The anxiety reward reflects the driver's satisfaction with the current SoC.

If EVs are the only flexible loads on RDN, we assume that all EVs indirectly interact with the global environment(RDN) through charging piles connected to aggregators. Thus the optimal power flow in \eqref{OPF obj} can be calculated by the management system of RDN. Then, the power change at time $t$ on the substation is calculated as
\begin{equation}
\label{power change}
g_t(a_{1,t},\cdots,a_{N,t})=f_{\text{obj},t}(P_{agg,t}) -f_{\text{obj},t}(0), 
\end{equation}
which we write as $g_t$ in the following for notation simplification. Here, $f_{\text{obj},t}$ in \eqref{OPF obj} is a function of $P_{agg_s,t}$ set, thus, $g_t$ is actually a function of $(a_{1,t},\cdots,a_{N,t})$ according to \eqref{agg consists of EV}. RDN will broadcast the power change $g_t$. Each EV receives power change $g_t$ from the global environment through communication. 
Then, the EV$i$ can quantify charging/discharging power except itself on RDN, and we regard it as grid reward, i.e.,
\begin{equation}
\label{grid reward}
r_{i,t}^{\text{g}}=\begin{cases}
-\max \left( g_t-g_{i,t},0 \right) &a_{i,t}>0,\\
\min \left( g_t-g_{i,t},0 \right) &a_{i,t}< 0 ,\\
0 &a_{i,t}=0 , \\
\end{cases} 
\end{equation}
with $g_{i,t}=g_t(0,\cdots,a_{i,t},\cdots,0)$. The grid reward reflects RDN load, i.e., the idle degree of RDN. Specifically, the $g_{i,t}$ represents the impact of EV$i$ charging power on RDN. 
The $g_t-g_{i,t}$ is the difference between the charging/discharging trends of the RDN and the charging/discharging power of an individual EV. If $g_t-g_{i,t}>0$ when $a_{i,t}>0$, some other EVs are charging besides EV$i$. EVs charging concentrated will burden RDN. So we penalize such action with a negative value. And if $g_t-g_{i,t}<0$ when $a_{i,t}<0$, EVs discharging concentrated will also cause an increase in node voltage, thereby affecting the RDN operation. Therefore, the discharging behavior of EV$i$ is discouraged in this case. When $a_{i,t}=0$, the EV$i$ neither charges nor discharges. Thus, the grid reward is set as 0. In our assumption, charging will impose a greater burden on RDN because EV’s charging power is larger than its discharging power.

The sum reward for EV$i$ at time $t$ combines the power reward, anxiety reward, and grid reward:
\begin{equation}
\label{sum reward}
r_{i,t}(s_{i,t},a_{i,t})=\lambda_{\text{p}} r_{i,t}^{\text{p}} + \lambda_{\text{a}} r_{i,t}^{\text{a}} + \lambda_{\text{g}} r_{i,t}^{\text{g}},
\end{equation}
where weight coefficients $\lambda_{\text{p}}$, $\lambda_{\text{a}}$ and $\lambda_{\text{g}}$ depend on sensitivity to electricity price, driver's anxiety, and RDN load. The setting of these coefficients depends on the drivers' habits and the structure of RDN. All these coefficients are non-negative.

\subsection{Problem Formulation}


The distribution network multi-EV G2V/V2G system aims to coordinate the EVs charging/discharging on RDN subjected to the physical model and the operational constraints. A global policy is learned to map the states of each agent to probability action space. The ultimate optimization objective is to obtain a global charging control policy $\pi$ to maximize the expected cumulative reward:
\[
\max_{\pi }\left( \frac{1}{N}\sum_{i=1}^N{\mathbb{E}_{a_{i,t}\sim \pi _i,s_{i,t+1}\sim \mathbb{P}_i}\left[ \sum_{t=1}^T{\gamma}^t\cdot r_{i,t}\left( s_{i,t},a_{i,t} \right) \right]} \right) ,
\]
where $\gamma \in \left[ 0,1 \right] $ reflects importance of future reward and $T$ is the horizon of EV charging control process. Note that the OPF-based RDN shields the detailed interaction between the charging/discharging power of each EV. As a result, each agent only requires local observation in our proposed environment, simplifying the problem.

\section{Methodology}
\label{sec::algorithm}

In this section, we first summarize the Soft Actor-Critic (SAC), used to control the charging behavior of a single EV agent. We then introduce FedSAC, a novel approach that combines SAC with Federated Learning to overcome data isolation issues in a privacy-preserving manner.

\subsection{Reinforcement Learning for Single EV Charging}

Reinforcement Learning enables an agent to learn through interactions with the environment and make near-optimal sequential decisions. SAC is one of the off-policy Reinforcement Learning frameworks based on maximum entropy to address the poor stability of model-free DRL methods. SAC aims to learn a more diverse policy by balancing exploration and exploitation with entropy regularization, which prevents the agent from getting stuck in local optima.
The experimental results demonstrate that SAC achieves state-of-the-art performance on various benchmark tasks. Therefore, we adopt it in this paper to make optimal decisions in EV charging control. 

The objective of SAC for EV$i$ is to determine its associated optimal local policy $\pi_i^{}$, which maximizes cumulative reward and policy entropy. This can be achieved through the following equation:
\begin{align}
\pi_{i} ^{*}=\underset{\pi}{\text{arg}\max}\,\,\mathbb{E}_{a_{i,t}\sim \pi \left( \cdot \mid s_{i,t} \right)}\bigg[ \sum_{t=0}^{T}&\gamma^t\Big( r_t\left( s_{i,t},a_{i,t} \right) \\\notag
&\qquad +\alpha_{i} \mathcal{H}\left( \pi \left( \cdot |s_{i,t} \right) \right) \Big) \bigg],
\end{align}
where $\mathcal{H}\left( \cdot \right)$ denotes the entropy regularization, and the $\alpha_{i}$ is a temperature coefficient that controls the weight between cumulative reward and policy entropy.

The soft Q-function in SAC for EV$i$ is defined as
\[
Q\left( s_{i,t},a_{i,t} \right) =r_t\left( s_{i,t},a_{i,t} \right) +\gamma \mathbb{E}_{s_{i,t+1}\sim \mathbb{P}_{i}}\left[ V\left( s_{i,t+1} \right) \right],
\]
where $V$ is a soft state value function approximated by the Q-function as 
\[
V\left( s_{i,t} \right) =\mathbb{E}_{a_{i,t}\sim \pi _{i}}\left[ Q\left( s_{i,t},a_{i,t} \right) -\alpha_i \log \left( \pi \left( a_{i,t}|s_{i,t} \right) \right) \right] .
\]
The critic is approximated by a Deep Neural Network (DNN) parametric over $\theta_i$, i.e., $Q_{\theta_{i}}(s_{i,t},a_{i,t})$. 
The soft Q-function is learned as a regression problem to minimize the soft Bellman residual $J_{Q}\left( \theta _{i} \right):=$
\begin{equation}
\label{critic loss}
\mathbb{E}_{\left( s_{i,t},a_{i,t} \right) \sim \mathcal{D}_{i}}\left[ \frac{1}{2} \left( Q_{\theta _{i}}\left(s_{i,t},a_{i,t}\right)-Q_{\hat{\theta}_{i}}\left(s_{i,t},a_{i,t}\right)\right)^2 \right] ,
\end{equation}
where
\[
Q_{\hat{\theta}_{i}}\left(s_{i,t},a_{i,t}\right)=r_t\left( s_{i,t},a_{i,t} \right) +  \gamma \cdot \mathbb{E}_{s_{i,t+1} \sim p} V_{\hat{\theta}_{i}}\left( s_{i,t+1} \right)
\]
sampling mini-batches from a replay buffer $\mathcal{D}_{i}$. The $V_{\hat{\theta}_{i}}$ is the estimated soft state value defined by a target DNN parametric over $\hat{\theta}_i$. The parameter $\hat{\theta}_{i}$  can be updated by moving average method $\hat{\theta}_{i}=\tau \theta _{i}+\left( 1-\tau \right) \hat{\theta}_{i}$.··

The policy function is also approximated by a DNN parametric over $\phi_{i}$. The policy network is updated by minimizing Kullback-Leibler (KL) divergence $J_{\pi}\left( \phi_i \right):=$
\begin{align}
\label{actor loss}
 \mathbb{E}_{\substack{s_{i,t}\sim \mathcal{D}_{i} \\ 
a_{i,t}\sim \pi _{\phi_{i}}}}\left[ \alpha_{i} \log \pi _{\phi_{i} }\left( a_{i,t}|s_{i,t} \right)-Q_{\theta _{i}}\left( s_{i,t},a_{i,t} \right) \right] .
\end{align}

Because the charging rate of EV is continuous in the charging control process, we set the policy $\pi_{\phi_{i}}$ as a Gaussian distribution, i.e.,
\begin{equation}
\label{Gaussian distribution}
\pi _{\phi_{i} }\left( a_{i,t}|s_{i,t} \right) =\frac{1}{\sqrt{2\pi \sigma}}\exp \left( -\frac{\left( a_{i,t}-\mu \right)^2}{2\sigma ^2} \right) ,
\end{equation}
where $\mu$ and $\sigma $ are outputs of the policy network, which are the mean and standard deviation of $s_{i,t}$, respectively. As $\mu$ and $\sigma$ are the function of $s_{i,t}$, we can rewrite them as $\mu=\mu_{\phi_{i}}(s_{i,t})$ and $\sigma=\sigma_{\phi_{i}}(s_{i,t})$. The policy network will output $\mu$ and $\sigma $ given the current state $s_{i,t}$. Then we randomly sample the current action $a_{i,t}$ from \eqref{Gaussian distribution}. 

To deploy backpropagation in the policy network, we introduce a reparameterization trick to rewrite the policy in \eqref{Gaussian distribution} as $f_{\phi_i}\left(\epsilon;s_{i,t}\right)= \mu _{\phi_i}\left( s_{i,t} \right) +\epsilon \sigma _{\phi_i}\left( s_{i,t} \right) $
with $\epsilon\sim\mathcal{N}\left( 0,1 \right) $. Specifically, the $\epsilon$ is sampled from the standard normal distribution and the $a_{i,t}= f_{\phi_i}\left(\epsilon;s_{i,t}\right)$ is then generated deterministically.
Thus the gradient of \eqref{actor loss} is approximated by 
\begin{align}
\label{actor gradient}
&\nabla_{\phi_{i}} \log \pi_{\phi_{i}}\left(a_{i,t} \mid s_{i,t}\right) +\Big(\nabla_{a_{i,t}} \log \pi_{\phi_{i}}\left(a_{i,t} \mid s_{i,t}\right)\\\notag
&\qquad\qquad-\nabla_{a_{i,t}} Q_{\theta_{i}^{k}}\left(s_{i,t}, a_{i,t}\right)\Big) \nabla_{\phi_{i}} f_{\phi_{i}}\left(\epsilon ; s_{i,t}\right).
\end{align}
The choice of temperature coefficient affects the effectiveness of the policy network. Thus we use an adjustment strategy to update $\alpha_i$ automatically by minimizing 
\begin{equation}
\label{temperature loss}
J\left( \alpha_i  \right) =\mathbb{E}_{\substack{
s_{i,t}\sim \mathcal{D}_{i}\\
a_{i,t}\sim \pi _{i}}}\left[ -\alpha_{i} \log \pi _{\phi_i}\left( a_{i,t}|s_{i,t} \right) -\alpha_i \mathcal{H} \right] ,
\end{equation}
where $\mathcal{H}$ is the desired minimum expected target entropy.

In practice, two critic networks $Q_{\theta_{i}^{k}}$ and two target networks $Q_{\hat{\theta}_{i}^{k}}$ where $k\in\{1,2\}$ are utilized to diminish the overestimation of Q-value. So the equation in \eqref{critic loss} and \eqref{actor loss} can be rewritten as 
\begin{align}
J_{Q}\left( \theta _{i}^{k} \right)=\mathbb{E}_{\left( s_{i,t},a_{i,t} \right) \sim \mathcal{D}_{i}}\left[ \frac{1}{2} \left( Q_{\theta _{i}^{k}}-\min_{k} Q_{\hat{\theta}_{i}^{k}}\right)^2 \right] , \\
 J_{\pi}\left( \phi_i \right)=\mathbb{E}_{\substack{s_{i,t}\sim \mathcal{D}_{i} \\ 
a_{i,t}\sim \pi _{\phi_{i}}}}\left[ \alpha_{i} \log \pi _{\phi_{i}}-\min_k Q_{\theta _{i}^{k}} \right] .
\end{align}

Algorithm~\ref{alg:SAC-single} presents charging control of EV$i$ based on SAC. Firstly, EV$i$ clears its experience replay buffer $D_i$ and initializes network parameters. Afterward, the EV$i$ trains until a stable charging strategy is obtained. During the environment step, the EV interacts with local and global environments to collect experience. In the global environment, the RDN calculates the power change $g_t$ at time $t$ through solving OPF in \eqref{OPF problem}. It sends $g_t$ and real-time electricity price $\xi_t$ to all subordinate nodes at fixed intervals. The EV receives $g_t$, and $\xi _t$ from RDN and then executes grid reward $r_{i,t}^{\text{g}}$ through \eqref{grid reward}. The EV$i$ subsequently calculates the power reward $r_{i,t}^{\text{p}}$ and anxiety reward $r_{i,t}^{\text{a}}$ through interaction with the local environment. The sum reward $r_t\left( s_{i,t}, a_{i,t} \right)$ is executed, and the obtained state is stored into $D_i$ for the local training. In the gradient step, the EV$i$ locally updates the parameters, i.e, $(\theta_{i}^{k},\phi_i,\alpha_i, \hat{\theta}_{i}^{k})$, by randomly sampling experience from $D_i$. Here, the $\delta_{Q}$, $\delta{\pi}$ and $\delta{\alpha}$ are the learning rates. Finally, the EV generates its charging control strategy.

\begin{algorithm}[htbp!]
\caption{Soft Actor-Critic for Charging Control of Single EV} 
\label{alg:SAC-single} 
\begin{algorithmic}[1]
\renewcommand{\algorithmicrequire}{ \textbf{Input:}}
\renewcommand{\algorithmicensure}{\textbf{Output:}}
\STATE
Initialize the replay buffer, actor network with $\phi_i$ and (target) critic network with ($\hat{\theta}_{i}^{k}$) $\theta _{i}^{k}$
\FOR{each training episode}
\FOR{each environment step}
\STATE
\textbf{Local Environment:}
\STATE
Sample action $a_{i,t}$ from $\pi_i$ given current state $s_{i,t}$
\STATE
Execute $a_{i,t}$ and obtain next state $s_{i,t+1}$
\STATE
\textbf{Global Environment:}
\STATE
Solve OPF in \eqref{OPF problem} and obtain $g_t$ through \eqref{power change}
\STATE
Broadcast the result $g_t$ on a communication link
\STATE 
Broadcast the real-time electricity price $\xi _t$ on a communication link
\STATE
\textbf{Local Environment:}
\STATE
Receive $g_t$ and execute $r_{i,t}^{\text{g}}$ through \eqref{grid reward}
\STATE
Calculate the power reward $r_{i,t}^{\text{p}}$ and anxiety reward $r_{i,t}^{\text{a}}$ through \eqref{power reward} and \eqref{anxiety reward}
\STATE
Obtain $r_t\left( s_{i,t},a_{i,t} \right)$ through \eqref{sum reward}
\STATE
Store the experience $(s_{i,t},a_{i,t},r_t,s_{i,t+1})$ into $\mathcal{D}_i$
\ENDFOR
\FOR{each gradient step}
\STATE
Update critic network parameters $\theta_{i}^{1}$ and $\theta_{i}^{2}$:
\STATE
$
\theta _{i}^{k}\gets \theta _{i}^{k}+\delta _Q\nabla _{\theta _{i}^{k}}J_{Q}\left( \theta _{i}^{k} \right) ,\ k\in \{1,2\}
$
\STATE
Update actor network parameter $\phi_i$:
\STATE
$
\phi _{i}\gets \phi _i+\delta_{\pi}\nabla _{\phi _i}J_{\pi}\left( \phi _i \right) 
$
\STATE
Update temperature parameter $\alpha_i$ :
\STATE
$
\alpha _{i}\gets \alpha _i+\delta _{\alpha}\nabla J\left( \alpha _i \right) 
$
\STATE
Update target critic network with moving average method:
\STATE
$\hat{\theta}_{i}^{k}=\tau \theta _{i}^{k}+\left( 1-\tau \right) \hat{\theta}_{i}^{k},\ k\in \{1,2\}$
\ENDFOR
\ENDFOR
\end{algorithmic}
\end{algorithm}

\subsection{FedSAC Framework for Multi-EV Charging}

The SAC mentioned above is shared among all agents, thus providing high autonomy for each EV's charging control. However, the limited diversity of charging experiences within individual EVs can weaken the generalization performance of the agent. Moreover, due to the privacy-sensitive nature of charging/discharging behavior, sharing data among different EVs may pose a risk. To overcome these limitations, FL is proposed to train a highly generalized agent while preserving privacy \cite{xu2021multiagent}.

Federated learning is a distributed machine learning framework for exchanging intermediate parameters free of privacy disclosure during training. It consists of two participants: various data-owned clients and the central server. The clients update the parameters of the local model based on decentralized data and upload updates to a central server. The central server aggregates the parameters from the clients to update the global model. The server then broadcasts the updated global model to all clients and updates their parameters. This process is repeated multiple times until the model converges.

The algorithm procedure of FedSAC is shown in Algorithm~\ref{alg:FedSAC}. At the beginning of the training process, the server builds a global SAC model with parameters $\theta_{\text{G}}^{k}$ and $\phi_{\text{G}}$. The $N$ clients train their own SAC models on local devices in parallel by following Algorithm~\ref{alg:SAC-single} from lines 4 to 29. The training on each client will iterate for $N_e$ episodes. The server receives the updated parameters from the clients, and aggregates them through $\theta_{\text{G}}^{k}=\frac{1}{N}\sum_{i=1}^N{\theta_{i}^{k}},\ k\in \{1,2\}$, and $\phi_{\text{G}}=\frac{1}{N}\sum_{i=1}^N{\phi_{i}}$. The new $\theta_{\text{G}}^{k}$ and $\phi _{\text{G}}$ replace the original parameters. The updated parameters are then broadcasted to the clients. The procedure of aggregation and broadcast repeats enough epochs until the global model converges.

\begin{algorithm}[htbp!]
\caption{Federated Soft Actor-Critic for Multi-EVs Charging Control} 
\label{alg:FedSAC} 
\begin{algorithmic}[1]
\renewcommand{\algorithmicrequire}{ \textbf{Input:}}
\renewcommand{\algorithmicensure}{\textbf{Output:}}
\STATE
Initialize the global actor network with $\phi_{\text{G}}$ and critic network with $\theta_{\text{G}}^k$
\FOR{global epoch}
\FOR{\textit{client $i$} from 1 to $N$ in parallel}
\STATE
Update $\theta_{i}^{k}$, $\phi_i$ and $\alpha_i$ locally: execute Algorithm \ref{alg:SAC-single} from lines 4 to 29 with $N_e$ looping episodes
\STATE
Upload $\theta_{i}^{k}$ and $\phi_i$ to the server
\ENDFOR
\STATE
Receive the parameters from each client
\STATE
Aggregate parameters of model through $\theta_{\text{G}}^{k}=\frac{1}{N}\sum_{i=1}^{N}{\theta _{i}^{k}},\ k\in \{1,2\}$ and $\phi _{\text{G}}=\frac{1}{N}\sum_{i=1}^{N}{\phi_{i}}$
\FOR{\textit{client $i$} from 1 to $N$ in parallel}
\STATE
Broadcast $\theta_{\text{G}}^{k}$ and $\phi_{\text{G}}$ to clients:  $\theta_{i}^{k} \gets \theta_{\text{G}}^k$, $\phi_{i} \gets \phi_{\text{G}}$
\ENDFOR
\ENDFOR
\ENSURE
$\phi_\mathrm{G}$
\end{algorithmic}
\end{algorithm}

\color{black}
\begin{figure}[htbp!]
\centering
\includegraphics[width=\linewidth]{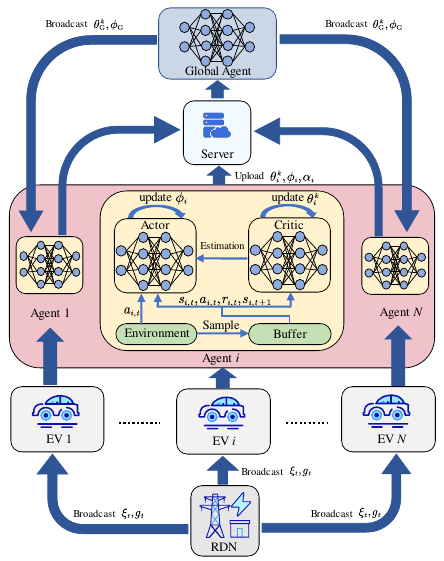}
\caption{The overall framework of FedSAC algorithm}
\label{Algorithm Framework}
\end{figure}

Fig.~\ref{Algorithm Framework} visualizes the overall framework of FedSAC applied in EV charging control process. The EV agent is a client responsible for local model training and communication with the server and the RDN. Local models are stored on individual EVs considering mobility and privacy. When an EV is plugged into the V2G charging pile, the embedded computer will take charge of the charging process. The onboard embedded computer generates a charging/discharging rate for the local EV for the next period at a specific time. The RDN manages its power flow by solving the OPF problem and executing $g_t$ at fixed intervals. 
Then, it broadcasts $g_t$ and real-time electricity price $\xi _t$ to all EVs (i.e., charging stations). The charging stations specify their unique price strategies according to $\xi _t$. Each EV calculates the individual rewards based on the global information from the RDN. After collecting enough rewards and states, such information will be used to update the parameters of SAC. The process is repeated enough times to generate a personalized charging policy for each EV. The EVs upload the updated parameters, and the server aggregates these parameters to generate a global model. The EVs download the global model to update the personalized charging policy. Training episodes repeat until the global model converges. Then a real-time charging control policy considering different driver behaviors is well-trained. As a standard federated learning framework, the FedSAC can train a highly generalized agent in a privacy-preserving manner.

\section{Experiment}
\label{sec::experiment}
In this section, we present several experiments to verify the performance of the proposed FedSAC applied to EV charging control. First, in Section~\ref{sec:Experiment Setting}, we describe the training settings and simulation setup. Then, in Section~\ref{sec:Training Performance}, we present the training performance of the proposed algorithm. In Section~\ref{sec:simulation results} and~\ref{sec:RDN simulation}, we conduct two simulations to verify the performance of charging control and charging power on RDN. Finally, in Section~\ref{sec:Comparative Evaluation}, we compare the performance of the proposed algorithm with different algorithms.

\subsection{Experiment Setting}
\label{sec:Experiment Setting}To demonstrate the effectiveness and performance of the proposed algorithm, we conducted our experiments on a platform with an \texttt{Intel Xeon Silver 4310 CPU @ 2.10GHz} $\times$ 12 and 1 \texttt{NVIDIA RTX4090 GPU}. All algorithms are implemented by \texttt{Python 3.8.15} with \texttt{Pytorch 1.7.1.} The hourly electricity price of weekdays between July.1, 2021, and June.30, 2022, from ISO New England \cite{ElectricityPrice} is used. We divide the extracted dataset into two parts. The first part includes the hourly electricity price of 100 weekdays used for simulation, and the second part is used to train the algorithm. The past 48 hours' electricity prices are input into the network to learn the trend of electricity prices.
\begin{figure}[htbp!]
\centerline{
\subfigure{\includegraphics{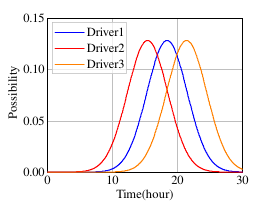}}
\subfigure{\includegraphics{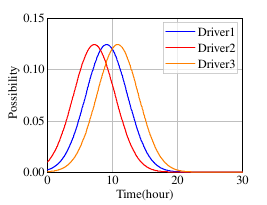}}
}
\centerline{
\subfigure{\includegraphics{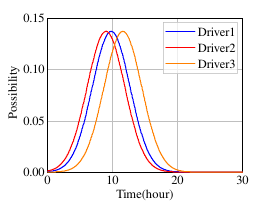}}
\subfigure{\includegraphics{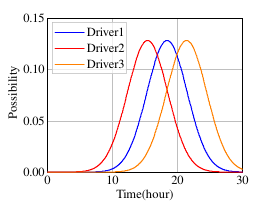}}
}
\caption{Statistical distribution of drivers' travel habits: Home Arrival(upper left); Home Departure(upper right); Office Arrival(lower left); (d) Office Departure(lower right).}
\label{driver habit}
\end{figure}

We study a thirty-car system with three kinds of travel and charging/discharging habits, as illustrated in Fig.~\ref{driver habit}(a)-(d). The curves in the figure indicate the probability of EV behavior. On weekdays, the kind-1 driver leaves home in the morning and arrives at the office after driving for a while. The driver finishes one day's work in the afternoon and departs the office. After driving for a period, the driver arrives home. We assume the EV is connected to the charging pile while not driving. The travel habit of kind-1 driver refers to the statistical analysis of the 2017 National Household Travel Survey (NHTS2017) \cite{TravelHabit}. Based on it, we assume the kind-2 driver prefers to return to the office early, and the kind-3 driver prefers to work overtime and return late. We sample the expected SoC $\beta_{i}^{1}$ at departure time $t_{i}^{d}$ for each EV driver from the ranges $[0.85,0.95]$, $[0.85,0.9]$ and $[0.9,0.95]$ respectively. The duration of three EVs is randomly distributed in $[1,4]$, $[1,2]$, and $[2,4]$. The ratio of three different kinds of EV drivers is 3:1:1. Each charging station sets a different price strategy to attract drivers, which means that each EV faces different electricity prices simultaneously. Despite the unit cost of electricity potentially varying among charging stations, they must still purchase electricity from the RDN. Therefore, the prices conform to the general trend of electricity prices within the grid. To simulate this situation, we randomly sample from [-10,10] and add it to the electricity price for every charging station at each timeslot. Additionally, we add a time offset of [-4, 4] to simulate any advance or delay in the electricity price trend. The EVs share some common parameter settings as well, including the initial SoC that is uniformly sampled from $[0,0.95]$, the shape parameter $\beta_{i}^{2}$ that is sampled from $\mathcal{N}(9,1^2)$ with bounds $[6,12]$, and the charging/discharging efficiency $\eta_c,\eta_d$ that is set to 0.98. Each EV battery's energy storage capacity $C_i$ is set to 0.03, meaning it can store 30 kW$\cdot$h of energy. We restrict the charging rate to $[-0.2,1.0]$ to simulate the imbalance between the charging and discharging rates. The weight coefficients in \eqref{sum reward} is set to $\lambda_{\text{p}}=9$, $\lambda_{\text{a}}=1$, $\lambda_{\text{g}}=100$ and  $\kappa _{\text{ta}}=36$, $\kappa _{\text{ra}}=16$.

Our research uses a 74-bus distribution network based on the IEEE 74-bus \cite{case74ds} to construct the global environment. We assume the presence of G2V/V2G DC charging stations on odd buses. Thirty EVs are connected to them separately when not driving. We assume that the charging piles can provide the maximum charging/discharging rate required by the EVs. In our research, the optimal power flow (OPF) problem is solved through the primal-dual interior point method \cite{InteriorPointMethod}.

\begin{figure}[htbp!]
\centerline{
\subfigure{\includegraphics{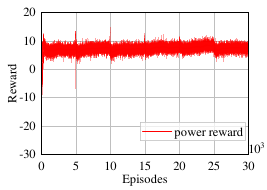}}
\subfigure{\includegraphics{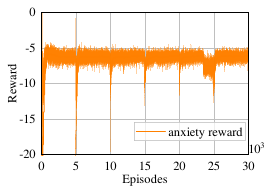}}
}
\centerline{
\subfigure{\includegraphics{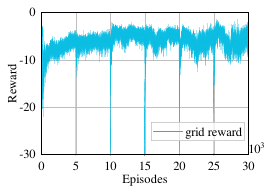}}
\subfigure{\includegraphics{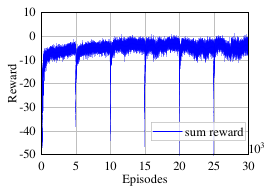}}
}
\caption{The aggregation reward curve during the training of FedSAC: power reward (upper left); anxiety reward (upper right); grid reward (lower left); sum reward (lower right).}
\label{reward curve}
\end{figure}

In the SAC framework, the critic network consists of three fully connected hidden layers with $128$ neurons each and rectified linear unit (ReLU) activation functions are applied after each hidden layer to bound the output. The input to the critic network concatenates the state $s_{i,t}$ and action $a_{i,t}$. A linear layer generates the final output Q-value. On the other hand, the actor network consists of four fully connected hidden layers with $\{128, 128, 128, 128\}$ neurons. The input to the actor network is the current state $s_{i,t}$, and two parallel linear layers generate the output mean ($\mu$) and standard deviation ($\sigma$). The size of reply buffer $\mathcal{D}_i$ is set to 10000, and the learning rate for the actor, critic, and temperature parameters are $10^{-4}$, $3\times 10^{-4}$ and $2\times 10^{-4}$, respectively. The batch size is 512, and the discount factor $\gamma$ is 0.99. The training process is separated into six rounds, each conducting $5000$ episodes. The FL mechanism is applied after every round.

\subsection{Training Performance}
\label{sec:Training Performance}

The reward curves of thirty agents during the training process are shown in Fig. \ref{reward curve}. The figure includes four kinds of reward curves: power reward, anxiety reward, grid reward, and sum reward. Each of them is the average of the reward curves of thirty EVs. The sum reward curve is the sum of the other three parts, which refer to \eqref{sum reward}. The shadowed regions represent the real training reward per episode, and the deep-colored curves illustrate smoothed episode reward curves. Although the natural average reward curves of the thirty agents exhibit significant fluctuations, the smoothed sum reward curves exhibit gradual growth and tend to stabilize eventually, indicating that our algorithm has learned a stable global policy. Since the FL mechanism averages the parameters of agents, the training reward would dramatically decrease during adjacent rounds, i.e., the reward curve drops at the 5000th, 10000th, etc, episode. 

Specifically, the algorithm converges rapidly in the first round. After applying the FL mechanism, the reward curve drops briefly at the beginning of the second round. The power and anxiety rewards tend to be even with those of the first round, and the trend repeats throughout the training process. Overall, the power reward and anxiety reward curves are relatively stable during federated learning. Besides, the grid reward converges quickly in the first two rounds and falls into local optimal convergence at the end of 2nd round. However, with the help of the FL mechanism, the grid reward overcomes this and rises to a higher level in the subsequent rounds. The grid reward tends to be stable after about the 20000th episode. This phenomenon illustrates that the FL mechanism helps break through the limitations of a single EV dataset, thus improving the generalization of the algorithm. The grid reward is related to the other EVs in the RDN, in other words, globally relevant, while power reward and anxiety reward reflect the personalized requirement of each EV. Therefore, the federal mechanism has a more significant effect on improving the grid reward. During the training process, each reward curve shows an upward trend, indicating that our algorithm considers every optimization objective, and the weight coefficients set achieves a balance between several objectives. The sum reward situation is similar to that of grid rewards, and the final stable trend indicates that our proposed algorithm has learned a generalized global policy. In summary, the FL mechanism would not harm the convergence process and could help isolated agents escape from the local optimum caused by limited charging experience. 

\subsection{Simulation of EV Charging Control}
\label{sec:simulation results}

\begin{figure}[htbp!]
\centerline{
\subfigure[Charging decision with different electricity prices]{\includegraphics[width=1.1\linewidth]{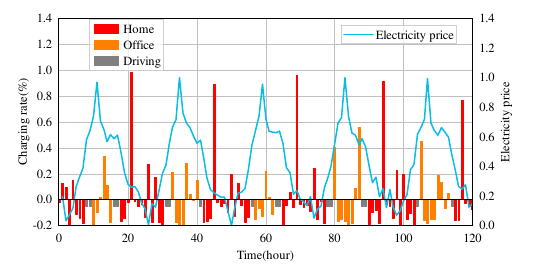}}
}
\centerline{
\subfigure[EV SoC at a different time]{\includegraphics[width=1.1\linewidth]{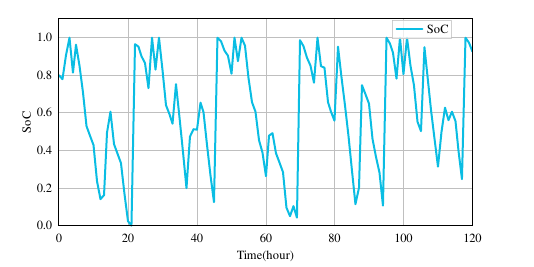}}
}
\caption{Simulation of EV charging control for Kind-1 Drivers in five weekdays.}
\label{simulation of EV1}
\end{figure}

\begin{figure}[t]
\centerline{
\subfigure[Charging decision with different electricity prices]{\includegraphics[width=1.1\linewidth]{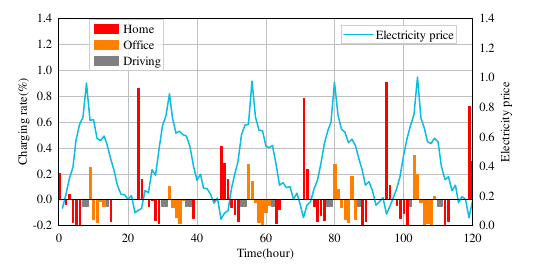}}
}
\centerline{
\subfigure[EV SoC at a different time]{\includegraphics[width=1.1\linewidth]{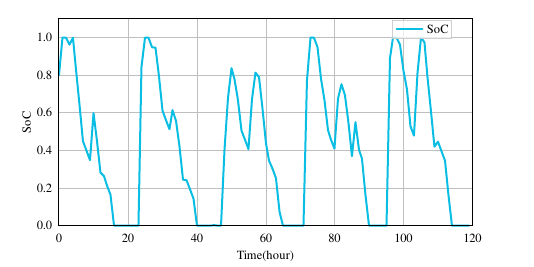}}
}
\caption{Simulation of EV charging control for Kind-2 Drivers in five weekdays.}
\label{simulation of EV2}
\end{figure}

\begin{figure}[t]
\centerline{
\subfigure[Charging decision with different electricity prices]{\includegraphics[width=1.1\linewidth]{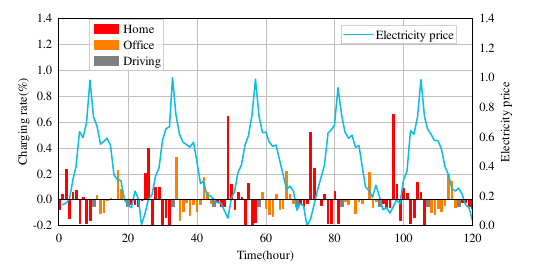}}
}
\centerline{
\subfigure[EV SoC at a different time ]{\includegraphics[width=1.1\linewidth]{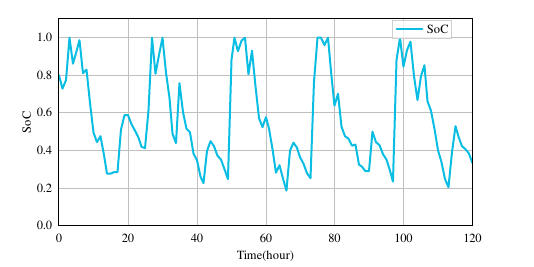}}
}
\caption{Simulation of EV charging control for Kind-3 Drivers in five weekdays.}
\label{simulation of EV3}
\end{figure}

We design a five-weekday uninterrupted trip to verify the performance of the charging control policy. In our simulation, EV drivers still follow the travel habits in Fig. \ref{driver habit}. We selected one of the three types of EV to observe the charging control strategy of each type of EV. When EV is in driving mode, the consumed energy is related to driving habits, driving distance, and environmental uncertainties (i.e., road conditions and weather). To simplify, we assume that the power consumption rate of each EV is 5\% per hour. The simulation results are shown in Fig. \ref{simulation of EV1}-\ref{simulation of EV3}. The colored columns of different lengths in Fig. \ref{simulation of EV1}-\ref{simulation of EV3}(a) represent charging/discharging rates in one hour at different locations. The positive rates indicate the EV is charging currently, while the negative one means the EV is discharging. The line reflects the electricity price trend, normalized and consistent throughout the RDN. The line in Fig. \ref{simulation of EV1}-\ref{simulation of EV3}(b) reflects the SoC change during the charging control process. 

As demonstrated in Fig. \ref{simulation of EV1}-\ref{simulation of EV3}(a), three EVs commonly prefer charging at low electricity prices and discharging at high prices, regardless of their location, as a means to minimize costs. The proposed method can reduce costs by employing dynamic electricity prices within the charging control process. In other words, the EVs can effectively regulate the sequential load distribution of RDN and further stabilize grid load under the charging decisions. Besides, when other EVs work in charging mode on RDN in a timeslot, the current EV would avoid charging along with them to prevent exorbitant instantaneous power. For instance, since electricity prices reach the first valley at approximately the 21st hour, three EVs opt to charge and reduce costs at this time. Considering the load pressure on the relevant distribution network, the EVs are charged in the order of EV1, EV2, and EV3. This result confirms that the proposed approach recommends dispersed charging in time from a global perspective. Likewise, node injection can potentially elevate the voltage at grid nodes. To counteract this, our algorithm regulates the decentralized discharging of EVs. The above phenomenon occurs because the grid reward effectively considers the current grid load. Fig. \ref{simulation of EV1}-\ref{simulation of EV3}(b) shows that the state of charge (SoC) remains no less than 0 throughout the simulation, indicating that the EV battery has ample power whenever needed for travel, thereby satisfying driver anxiety. The simulations confirm that the suggested algorithm enhances charging control policies, leading to lower charging expenses, reduced driver anxiety, and reduced RDN load. Moreover, this algorithm is suitable for all EVs.


\subsection{Power Load Performance under the Grid Reward}
\label{sec:RDN simulation}

The fluctuation in electricity prices exhibits a positive relationship with electricity demand. Under our policy, EVs prefer charging at times of low demand and discharging when demand is high, enabling them to play a novel role in electricity transactions. Under our assumption, the charging power of EVs is typically high, and simultaneous charging of all EVs could significantly augment the burden on the RDN, resulting in a rise in the load curve at the original valley and creating uncertainty in the power grid. Similarly, discharging also has the potential to introduce uncertainty, although the discharging power is typically lower than that of charging. Concentrated discharging can lead to an increase in node voltage, which further contributes to uncertainty. Such charging decisions would contravene the original purpose of V2G technologies. The simulation results in \ref{sec:simulation results} have proved that the three EVs would charge dispersively at the electricity peak. In this section, we design an ablation experiment to further assess the performance of the proposed approach from the perspective of the RDN.

\begin{figure*}[htbp!]
\centerline{
\subfigure[SAC without grid reward]{
\includegraphics{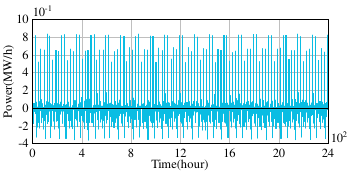}
}
\subfigure[SAC-MARL with grid reward]{
\includegraphics{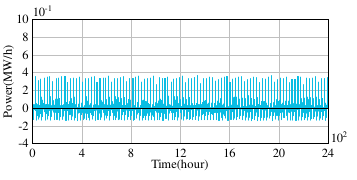}
}
\subfigure[FedSAC with grid reward]{
\includegraphics{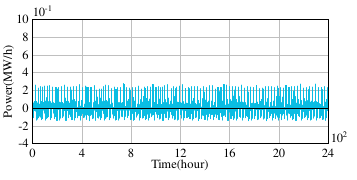}
}
}
\centerline{
\subfigure[DDPG-MARL with grid reward]{
\includegraphics{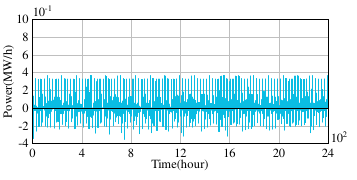}
}
\subfigure[TD3-MARL with grid reward]{
\includegraphics{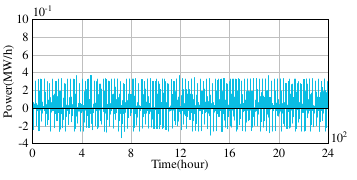}
}
}
\caption{Charging/Discharging power of thirty EVs on RDN.}
\label{Charging power change}
\end{figure*}

We design a SAC-based algorithm (refer to \cite{yanlingfang}) without the grid reward from the global environment and a multi-agent SAC (SAC-MARL) algorithm with grid reward. The other settings keep consistent with the proposed approach. We conduct a 100-weekday charging control simulation for both algorithms and get hourly charging/discharging action $a_{i,t}$ for all EVs. To better demonstrate the charging load on the RDN, we calculate the power change $g_t$ according to \eqref{power change}. The hourly power changes of the two algorithms are displayed in Fig. \ref{Charging power change}(a)-(b), with the colored bars representing the overall power the thirty-EV system is charging/discharging. Specifically, the charging/discharging power on the RDN in Fig. \ref{Charging power change}(a) fluctuates considerably with time under the control of SAC without grid reward. Conversely, the charging/discharging power on the RDN in Fig. \ref{Charging power change}(b) exhibits a comparatively smoother trend under the control of SAC-MARL, with a significant reduction in extreme power levels.

To quantify the power change resulting from charging/discharging on RDN during the simulation process, the standard deviation of power change ($\sigma_g$) is compared. The variance of power changes in the distribution network quantifies their degree of dispersion, where smaller power changes result in reduced uncertainty factors in the power grid. Specifically, the standard deviations of the SAC-MARL and SAC are 0.0919 and 0.5163, respectively. The SAC-MARL has reduced the standard deviation by 82.2\%, compared to the SAC algorithm without the grid reward. These results indicate that the algorithm with our proposed grid reward provides a more stable power curve. Hence, the algorithm with our proposed grid reward can regulate the charging behavior of each EV to apply a stabler charging load on the power grid.

\subsection{Comparative Evaluation}
\label{sec:Comparative Evaluation}

In this section, we evaluate the effectiveness of the FL mechanism by comparing it with several benchmarks as follows: 

\subsubsection{SAC-MADRL}

Multi-agent SAC takes the RDN as the global environment and considers the grid reward. The parameters setting and model structure are identical to that of FedSAC. SAC-MARL reduces the overhead of aggregation and broadcasting compared to FedSAC.

\subsubsection{DDPG-MADRL}

Deep deterministic policy gradient (DDPG) is an off-policy algorithm for continuous action spaces, including two actor and two critic networks. The multi-agent version introduces the global environment and considers the grid reward. The parameters setting and model structure are the same as that of FedSAC.

\subsubsection{TD3-MADRL}

Twin delayed deep deterministic policy gradient (TD3) is an extension of the DDPG. TD3 employs the clipped double-Q learning to prevent overestimation and stabilize learning. The multi-agent version introduces the global environment and considers the grid reward. The parameters setting keeps consistent with FedSAC.

\begin{figure}[t]
\centerline{
\subfigure{\includegraphics[width=1.1\linewidth]{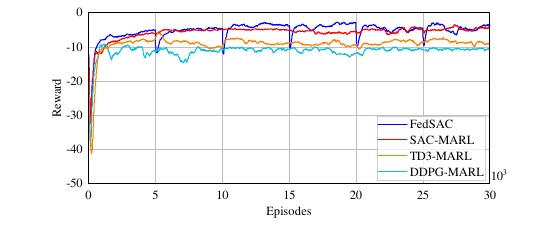}}
}
\caption{Comparisons of training process.}
\label{reward comparison}
\end{figure}

We compare the above algorithms from two dimensions: the
convergence and generalization. The convergence of agents based on various algorithms is demonstrated with average sum reward curves during training. The smooth reward curves for the thirty agents are shown in Fig. \ref{reward comparison}. In addition, we choose testing rewards and standard deviation of power change during the simulation as metrics to compare the generalization of thirty EVs. The test reward evaluates the performance of the simulated EV charging control strategies, with a higher value indicating better performance. The meanings of the variance of power grid changes have been elaborated in Section~\ref{sec:RDN simulation}. The simulation settings are consistent with those described in the preceding section. The testing rewards and average/standard deviation of power change are listed in Table~\ref{table:simulation performance}. The power changes of different algorithms on RDN are shown in Fig.~\ref{Charging power change}.

Fig.~\ref{reward comparison} illustrates the convergence process of thirty EVs. All the algorithms achieve a well-converged state under the same number of training episodes.  Our proposed FedSAC algorithm, which applies FL mechanisms, achieves the highest sum reward and is the best-performing algorithm. While the SAC-MARL algorithm also performs well, its limitation of isolated data makes its performance slightly inferior to that of FedSAC. However, the reward curve of FedSAC exhibits more volatility than SAC-MARL, demonstrating that FedSAC is slightly inferior in terms of stability. This is attributed to the non-uniformity and heterogeneity of data in distributed environments. TD3-MARL and DDPG-MADRL, which lack exploration of the environment, are far inferior to the SAC-based algorithms. Due to the optimization of the Q-function by the TD3 algorithm, its performance is still significantly better than that of DDPG-MARL.

Table~\ref{table:simulation performance} compares the generalization of different algorithms, and Fig.~\ref{Charging power change} compares the charging/discharging impact on RDN. We compare the standard deviation of power change and their decline ratio compared to SAC defined in Section~\ref{sec:Comparative Evaluation}. The bold values in the table represent the best test results among all algorithms. As shown in Table~\ref{table:Cumulative rewards}, the average test reward of thirty EV agents under FedSAC is -3720, which surpasses other comparative algorithms. Furthermore, the $\sigma_g$ of thirty EV agents trained by FedSAC is 8.63 in Table. \ref{table:RDN indicators}, indicating the best performance among all algorithms. Therefore, the charging/discharging load on RDN (i.e., Fig.~\ref{Charging power change}(c)) controlled by FedSAC is more stable than that of other algorithms. On the other hand, the DDPG-MADRL performs the worst on both average reward and standard deviation of power change, which is consistent with the training analysis. In addition, the average reward and $\sigma_g$ of SAC-MADRL are -4412 and 9.19, respectively, which are the closest to the optimal values. This phenomenon clarifies that the FL mechanism can improve the generalization of SAC. Since the FedSAC outperforms comparative algorithms in test rewards and standard deviation, FedSAC has the best generalization effect.

The results above clarify that the introduction of the FL mechanism dramatically improves EV agents' convergence effect and generalization ability. In summary, the comparisons demonstrate that the proposed FedSAC outperforms existing algorithms in EV charging control.

\begin{table}[htbp!]
\caption{Performance of different algorithms during the simulation}
\label{table:simulation performance}
\centering
\subtable[Cumulative average test rewards of different algorithms]{
\setlength{\tabcolsep}{1.6pt}
\begin{tabular}{ccccc}
\toprule    
 & sum reward & power reward & anxiety reward & grid reward \\
\midrule
DDPG-MARL & -5855 & -211 & -4136 & -1508 \\[0.1cm]
TD3-MARL & -4763 & 396 & -3698 & -1461 \\[0.1cm]
SAC-MARL & -4412 & 519 & -3790 & -1141 \\[0.1cm]
FedSAC & \textbf{-3720} & 550 & -3227 & -1043 \\
\bottomrule
\end{tabular}
\label{table:Cumulative rewards}
}
\smallskip
\subtable[Standard deviation of power change and their decline ratio]{        
\setlength{\tabcolsep}{1.2pt}
\begin{tabular}{cccccc}
\toprule    
& FedSAC & SAC-MADRL & TD3-MADRL & DDPG-MADRL & SAC \\
\midrule
$\sigma_g$(1e-2) & \textbf{8.63} & 9.19 & 12.29 & 12.41 & 51.63 \\[0.16cm]
decline & \textbf{83.28\%} & 82.2\% & 76.2\% & 75.96\% & 0\% \\
\bottomrule
\end{tabular}
\label{table:RDN indicators}
}
\end{table}




\section{Conclusion}
\label{sec::conclusion}

This paper proposes a federated deep reinforcement learning algorithm for the multi-EV charging control on an RDN. A smart OPF-based RDN is built first, which includes numerous isolated EVs connected to it, and an agent is used to manage EV charging control. A mathematical model is proposed to denote the RDN idle degree. The EV charging control problem is formulated as an MDP considering V2G profits, RDN idle degree, and driver's anxiety. Firstly, the individual optimal charging control strategy of each EV is learned by SAC. Then the FL mechanism is introduced to generate a global model by aggregating the local parameters from each agent. The driver's privacy is guaranteed during training. The case studies based on real-world data demonstrate that the proposed FedSAC can learn a stable cooperative charging control strategy and encourages decentralized EV charging, significantly reducing the load fluctuations on RDN. Besides, the comparisons clarify that the proposed algorithm performs better on convergence effect and generalization ability.

In this paper, the proposed FedSAC has slight shortcomings in terms of stability. Some potential methods such as weighted FedAVG and local training optimization can be tried to address such issues. Besides, the mobility of EVs is not considered because their stochastic nature makes it challenging to model. The future work will investigate the modeling approach to EV mobility and consider how to introduce it into EV charging control.

\bibliographystyle{IEEEtran}
\bibliography{references}

\end{document}